# Nanoscale charge accumulation and its effect on carrier dynamics in tri-cation perovskite structures

*David Toth[1,2], Bekele Hailegnaw[3], Filipe Richheimer[4], Sebastian Wood[4], Fernando A. Castro[4], Ferry Kienberger[1], Markus C. Scharber[3] and Georg Gramse[1,2]\**

1. Keysight Technologies GmbH, Linz, Austria

2. Applied Experimental Biophysics, Johannes Kepler University, Linz, Austria

3. Linz Institute for Organic Solar Cells (LIOS), Johannes Kepler University, Linz, Austria

4. National Physical Laboratory, Teddington, Middlesex, TW11 0LW, United Kingdom

\* Corresponding author: georg.gramse@jku.at

**Abstract**

Nanoscale investigations by scanning probe microscopy have provided major contributions to the rapid development of organic-inorganic halide perovskites (OIHP) as optoelectronic devices. Further improvement of device level properties requires a deeper understanding of the performance-limiting mechanisms such as ion migration, phase segregation and their effects on charge extraction both at the nano- and macroscale. Here, we have studied the dynamic electrical response of $Cs_{0.05}(FA_{0.83}-MA_{0.17})_{0.95}PbI_{3-x}Br_x$ perovskite structures by employing conventional and microsecond time-resolved Kelvin probe force microscopy (KPFM). Our results indicate strong negative charge carrier trapping upon illumination and very slow (>1s) relaxation of charges at the grain boundaries. The fast electronic recombination and transport dynamics on the microsecond scale probed by time-resolved KPFM show diffusion of charge carriers towards grain boundaries and indicate locally higher recombination rates due to intrinsic spatial heterogeneity. The nanoscale electrostatic effects revealed are summarized in a collective model for mixed-halide CsFAMA. Results on multilayer solar cell structures draw direct relations between nanoscale ionic transport, electron accumulation, recombination properties and the final device performance. Our findings extend the current understanding of complex charge carrier dynamics in stable multi-cation OIHP structures.

**Introduction**

In the last decade, organic-inorganic halide perovskites (OIHP) have received tremendous attention for optoelectronic applications, which has triggered rapid scientific progress in this field [1,2]. In particular, single junction OIHP photovoltaic devices have reached efficiencies of ~25%, open circuit voltages >1.2 V and short circuit currents >20 mA [3,4]. Further advancement towards the Shockley-Queisser limit requires a comprehensive micro- and macroscale understanding of performance-limiting effects, such as ionic migration and phase segregation in OIHPs. Defect formation via these processes can lead to the creation of non-radiative recombination pathways that result in spatially inhomogeneous recombination times and energetically preferred recombination sites [5–7]. In typical I-Br mixed-halide perovskites, iodide and/or bromide rich phases tend to form under prolonged illumination [8,9]. Confocal photoluminescence (PL) measurements suggest that these phases form at grain boundaries (GB), where current-voltage curve hysteresis is shown to be increased, further indicating ionic displacement and trap site formation [6,10–12]. On the other hand, Li et al. have observed higher conductivity at GBs via conductive atomic force microscopy arguing that GBs act as conduction pathways rather than recombination sites [13]. Cesium-incorporated perovskites offer an attractive solution as they have been shown to constrain effects such as phase segregation while maintaining high efficiencies [14,15]. This puts them at the focus of device optimization research. As an example, tri-cation mixed-halide perovskite inverted structures utilizing formamidinium (FA), methylammonium (MA) and Cs cations were reported with increased fill factor, high power conversion efficiencies and improved charge extraction for a two layered electron transport material (ETM) arrangement. The cells were prepared as ITO/NiO$_x$/perovskite/PCBM/TiO$_x$/Al architectures, where ITO stands for indium tin oxide and PCBM for [6,6]-phenyl-C61 butyric acid methyl ester. The $Cs_{0.05}(FA_{0.83}\text{-}MA_{0.17})_{0.95}PbI_{3-x}Br_x$ devices with $TiO_x$ layer on top of PCBM have shown lower charge carrier extraction resistance via impedance spectroscopy as well as a reduced extent of charge trapping through PL decay measurements compared to devices without the $TiO_x$ layer [16]. Relating these macroscale differences to nanoscale electrical measurements can yield a deeper understanding of the underlying physical behavior.

As recent developments of advanced scanning probe microscopy (SPM) techniques now allow for the characterization of electrically and optically active materials with high temporal resolution, they offer insight into the spatial variation of charge carrier dynamics on multiple timescales [17–20]. These approaches enable non-invasive imaging down to picosecond temporal resolution [21,22]. Improvement of the time resolution has been achieved by various means, such as recording the cantilever deflection read-out signal with ultra-high sampling rate as well as by overcoming the averaging limitation by applying high frequency pulses in excitation and detection [23–26]. The latter so-called pump-probe based approaches have been used in optical spectroscopy for decades and recently have been applied to SPM in order to resolve the local decay of the surface photovoltage in organic donor-acceptor blends, contact potential difference (CPD) changes in an organic transistor structure with microsecond temporal resolution, and picosecond relaxations of photovoltage in low temperature grown GaAs [27,28]. Limitations and possible artefacts of this approach have also been investigated by a number of groups [27,29]. Murawski et al. reported on a dual-closed loop system that removes crosstalk artefacts in the topography channel originating from an incorrect voltage applied by the KPFM feedback loop [28]. We overcome this artefact by detecting the first harmonic KPFM signal, $F_\omega$ directly in open-loop measurements, however, we note that in this case additionally the second harmonic of the electrostatic force, $F_{2\omega}$, has to be considered to reveal the transient CPD change.

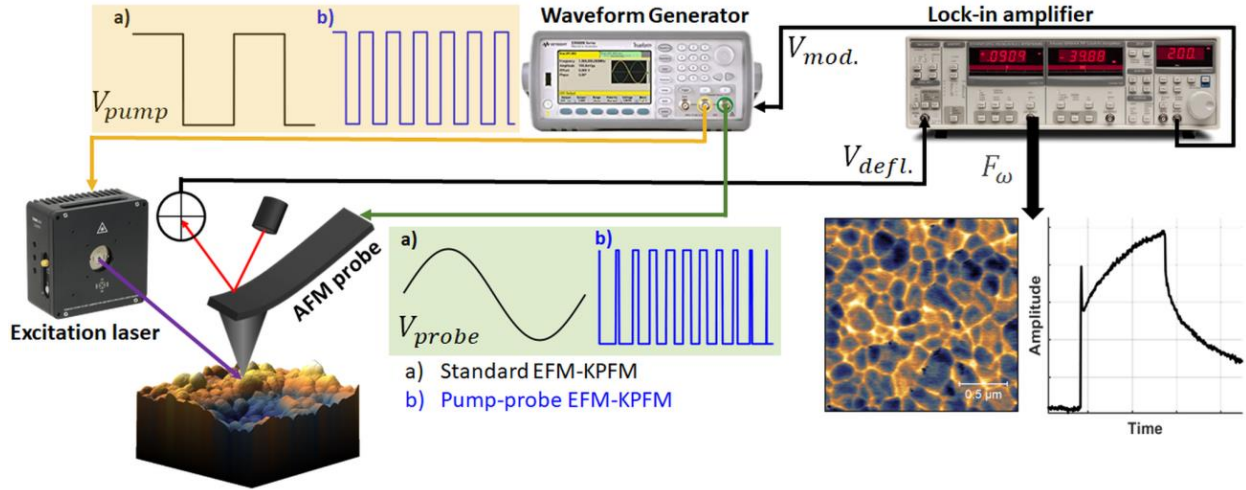

**Figure 1.** Connection scheme for (a) conventional and (b) pump-probe Kelvin probe force microscopy measurements. The electrically excited AFM probe is used to characterize the perovskite electrostatic response under illumination.

In order to shed light on the nanoscale spatial differences in the ionic and electronic dynamics of triple-cation mixed halide perovskite photovoltaic structures we employed conventional and pump-probe implementations of KPFM, respectively. The experimental setup is shown in **Figure 1** and is detailed in the Materials and Methods section. We performed measurements using two different light sources for excitation of the sample: a 633 nm CW laser coupled through an electro-optical modulator, and a directly modulatable 405 nm CW laser diode. The AC voltage applied to the tip was supplied externally and the demodulation of the detector signal was achieved with an external lock-in amplifier.

**Results and Discussion**

Upon optical excitation the electrostatic force response of a pristine CsFAMAPbBrI perovskite evolves in a complex way at timescales covering several orders of magnitude. This is shown in **Figure 2** where we display a typically measured $F_\omega$ response to a ~2.35 W/cm² 633 nm laser pulse with the dashed lines representing mono-exponential fits to separate features of the curve. Upon illumination electron-hole pairs are generated with charge carriers accumulating on each side of the perovskite depending on the sign of the voltage applied to the tip and the built-in electric field (step I). The electron mobility was shown to be in the range of 1-100 cm²V⁻¹s⁻¹ with characteristic lifetimes of nanoseconds to microseconds for OIHP by techniques such as terahertz electrical conductance [3]. Dynamics of this step are too fast to be resolved by standard electrostatic force microscopy techniques. Following the electron-hole pair generation, a process takes place on the scale of hundreds of milliseconds (step II). We interpret the relaxation of this overshoot as halide ion motion, as the timescales $\tau_{II}$ = 0.23 ± 0.05 s found in our measurements match with reported mobilities in the range of 10⁻⁷ cm² V⁻¹ s⁻¹ and a displacement rate of ~ 100 nm/10 ms [3,30,31].

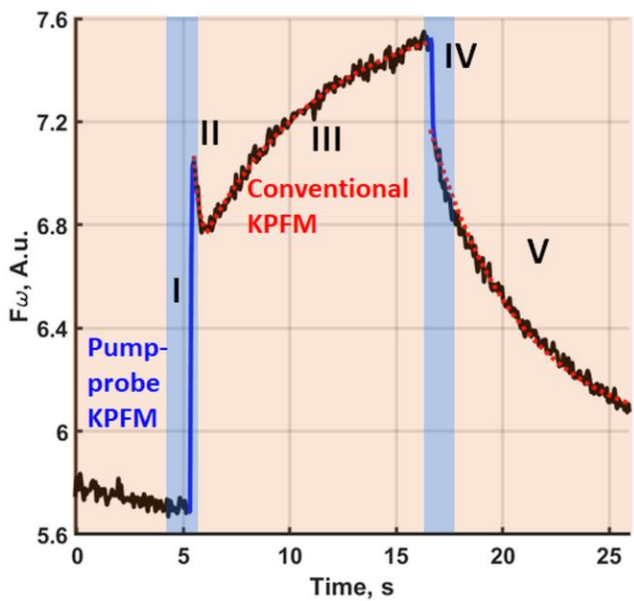

**Figure 2.** First harmonic of the electrostatic force response to a single light pulse (10.5 W/cm$^2$, 633 nm). The distinct dynamical processes are numbered I-V. Dashed lines indicate mono-exponential fit ($f(x) = y_0 + Ae^{(x/\tau)}$) of the separate features of the response. Processes in the red region in the second-millisecond range can be accessed by standard KPFM. Fast microsecond dynamics in the blue regions were measured by pump-probe KPFM.

Under illumination an additional process (step III) was recorded which reached a steady value only over the course of >10 s ($\tau_{III}$ = 5.9 ± 0.2 s). The dynamics of this process are much slower than the previously described halide ion/vacancy transport. A comparison of the time-evolution of confocal PL spectra recorded at various intensities (See Supporting Information) excludes the possibility that this behavior corresponds with phase segregation. **Figure S1A** shows that the PL spectrum measured with 17.5 W/cm$^2$ excitation is stable over a period of 90 s with the spectral peak at ~757 nm. In **Figure S1B** we show that the PL peak spectral position remains stable over time for low excitation intensities. Only above around 5.6 kW/cm$^2$ we do resolve the characteristic spectral shift of mixed halide perovskite phase segregation on the timescale relevant to the KPFM measurements in this work [32]. This suggests that any light-induced effect of phase segregation is negligible at the light intensity used here and that step III is of a different origin. Previous studies on cross-sectional samples have attributed contact potential difference ($V_{cpd}$) changes with similar dynamics to those observed here to charge accumulation and space charge formation [20,30]. Such a process would also result in a change of the $F_{2\omega}$ component, which is related to $dC/dz$, the capacitance gradient with respect to the tip-sample distance, z. By probing the change in the force component upon low power illumination (**Figure S2**), we demonstrate that the dynamics of step III result from a capacitive charging process.

After turning off the illumination, we first observe a fast recombination of electron-hole pairs at a time scale of nanoseconds to microseconds (step IV) and then a relaxation takes place (step V) with similar dynamics as the charge accumulation in step III ($\tau_V = 4.24 \pm 0.19$). The slow relaxation indicates charge trapping as the origin. Although a slow process can be seen at steps III and V in the first and the second harmonic signal, a change of work function is also expected upon accumulation or relaxation of electronic charge carriers.

**Charge accumulation and ionic processes.** To investigate whether the proposed charge accumulation and relaxation in step III and V results in a work function change we employed conventional KPFM mapping prior to and post illumination. **Figure 3A**, shows the measured topography of the CsFAMAPbBrI sample deposited on ITO. The image shows varying grain sizes with 100-500 nm diameter and an overall mean roughness of ~15 nm. The KPFM scan taken at dark conditions (**Figure 3B**) exhibits an average $V_{cpd}$ ~-160 mV with respect to the Au coated tip. The contrast between grains and grain boundaries indicates a larger work function at the GBs and a downward band bending towards the GBs. To probe the contribution of trapped charge carriers to the slow accumulation and relaxation processes we scanned the same area (within a few seconds) after an illumination period of ~7 minutes with a confocal beam at an intensity of 1.05 kW/cm$^2$. **Figure 3C**, shows the KPFM image taken after the illumination where mobile electron-hole pairs have completely recombined and the average sample potential changes to $V_{cpd}$~85 mV with respect to the probe. From the difference of the two maps we deduce the contribution of the trapped charges which results in the mean positive shift of $\Delta V_{cpd}$ = 243 mV. The positive shift of the $V_{cpd}$ is attributed to either a decrease of the sample work function or an accumulation of negative charges at the surface. In **Figure 3D**, we show the corresponding $\Delta V_{cpd}$ map overlaid on the surface topography where a larger $\Delta V_{cpd}$ can be seen at the GBs. To compare the overall shift in work function for different areas of the sample we prepared a mask of the grain boundaries from the topography image (shown in **Figure S3**) and used this to separate datapoints at grains and grain boundaries. This type of analysis indicates that the light-induced change in $V_{cpd}$ at the GBs (249 mV) is higher than that observed on the grains (237 mV) (**Figure 3E**). The observed downward band bending towards GBs in the dark KPFM images would facilitate the local accumulation of electrons. As the second map was taken within seconds after the illumination period, the results are confirmation that the observed contrast arises from trapped negative charges. Additionally, as the overlaid map exhibits a continuous change of the $\Delta V_{cpd}$ over time in dark conditions during the measurement in the slow scanning direction, this change can be directly related to step V already described in **Figure 2**. Cesium-incorporated perovskites have shown enhanced stability against ionic motion due to variation of lattice parameters across the structure [14,15]. Although, the stability of the PL peak position shown in **Figure S1** indicates reduced ionic movement, the contribution of ionic charge carriers at the nanoscale cannot be completely excluded from the observed effect.

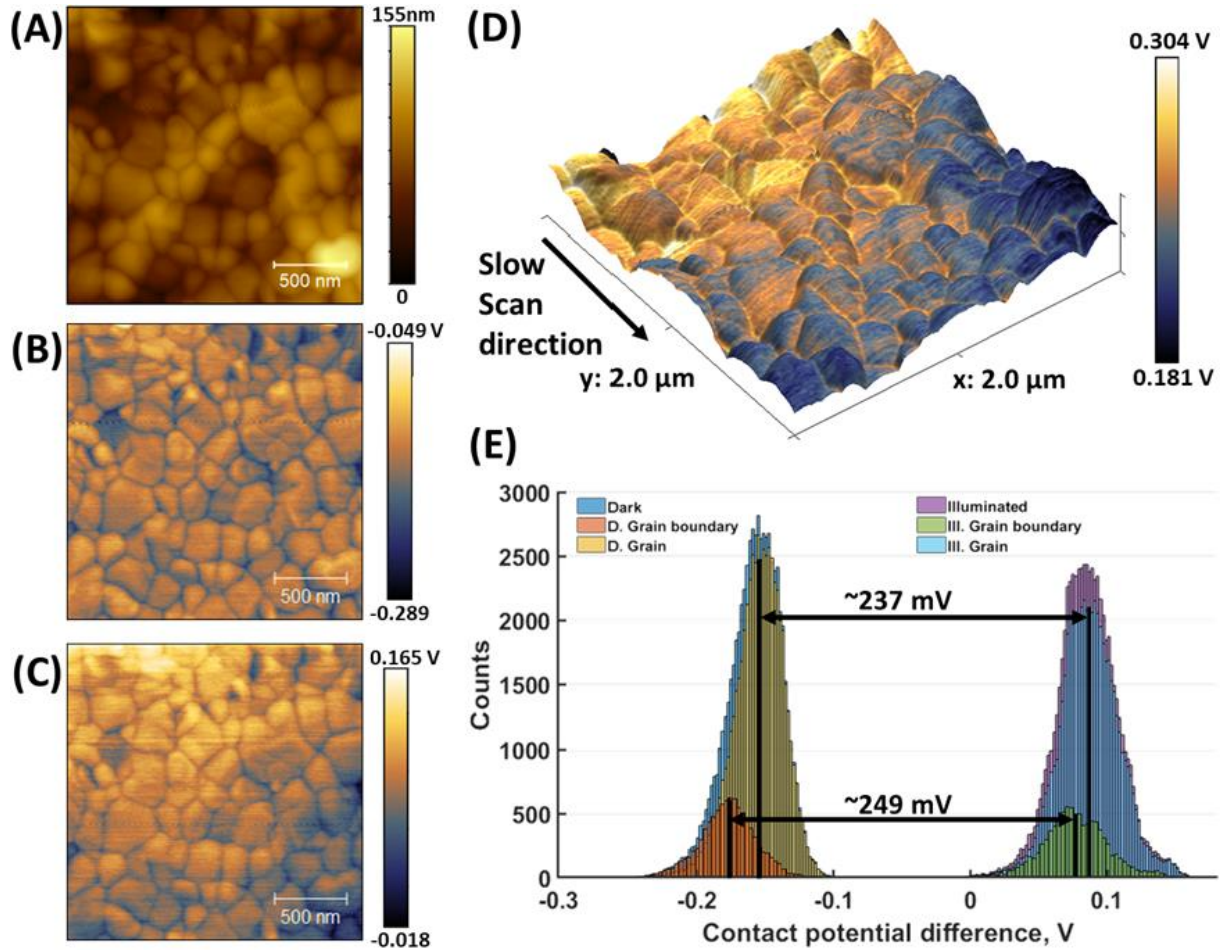

**Figure 3.** Initial dark and relaxed KPFM analysis of CsFAMAPbBrI. (A) Topography channel (B) KPFM image prior to light pulse (C) KPFM image seconds after the light pulse (D) 3D topography overlaid with the calculated $\Delta V_{cpd}$ map. (E) Histograms of before and after $V_{cpd}$ maps separated into grain and grain boundary responses. The double arrows indicate the difference between the mean value of the distributions.

**Open-loop pump-probe KPFM.** To further investigate the origin of the charge accumulation at the GBs, contribution of electronic and ionic charges has to be separated. While the local dynamics of the trapped charges and ions are slow enough to be measured by standard KPFM, the much faster electronic charge carrier recombination and diffusion requires a different approach. Here we employed open-loop pump-probe KPFM (pp-KPFM) for this. Briefly, in open-loop pp-KPFM two high frequency pulse signals with changing phase difference are applied to probe and light source, respectively, in order to resolve changes in the $F_\omega$ signal with high temporal resolution. In our approach, we used modulated laser pulses as the pump signal and a sinusoidally modulated pulse waveform applied to the tip as the probe signal and recorded $F_\omega$ as described in the Materials and Methods section. Additional details of this approach can be found in the Supporting Information. Pulse frequencies in the range of 50-150 kHz were selected to enable monitoring of microsecond timescale effects whilst still being modulatable by a sinusoidal waveform at the cantilever resonance frequency to improve sensitivity of detection. In **Figure S4A** we present a typical open-loop pp-KPFM response from the perovskite layer and the dependence of the extracted time constants on probe pulse-width.

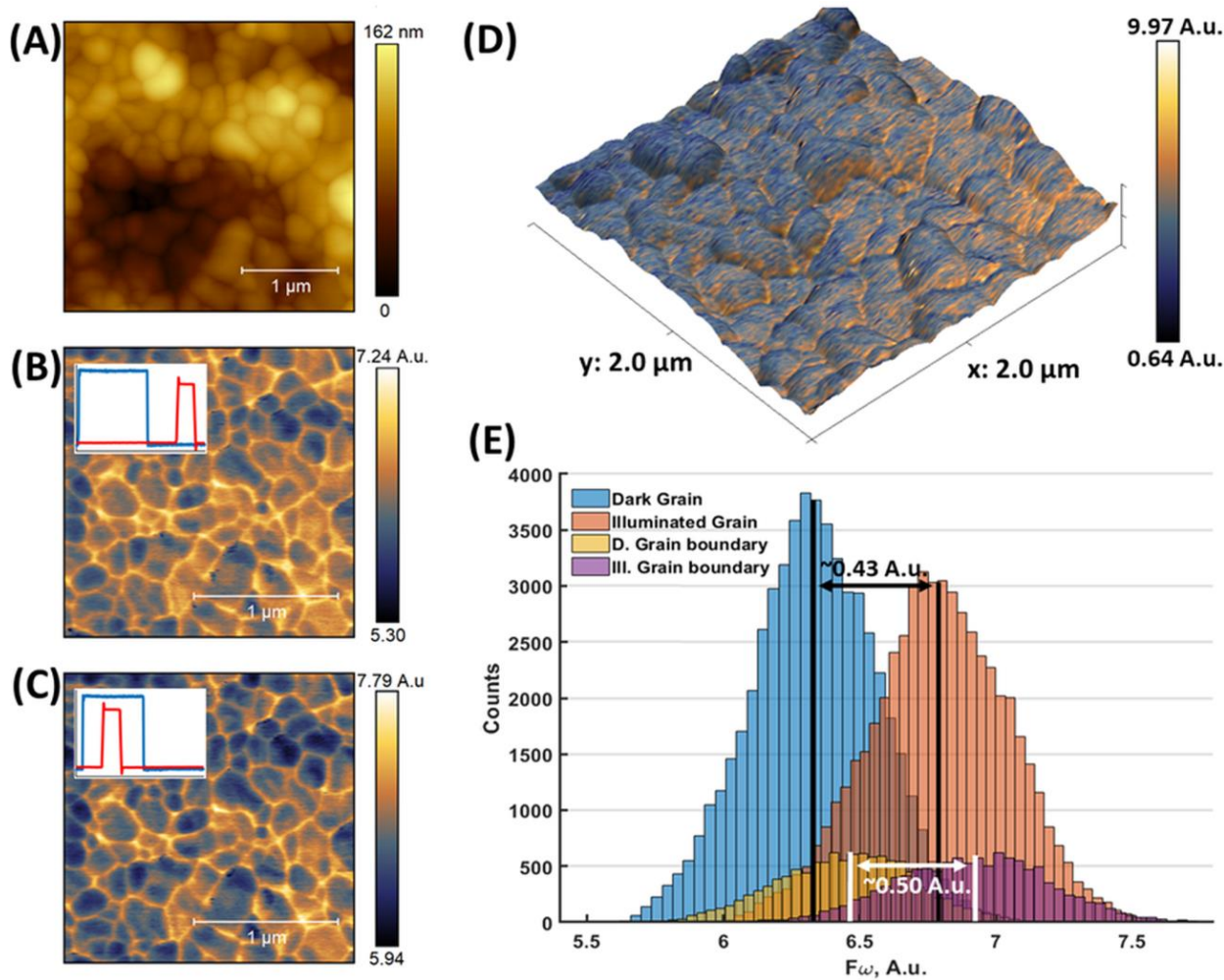

**Figure 4.** Pump-probe KPFM maps of mixed-halide CsFAMA at 150 kHz pulse frequencies. (A) Topography channel (B), (C) respectively $F_{\omega,off}$ (in the dark) and $F_{\omega,on}$ (under illumination) response at the pulse positions shown in the inset (where blue indicates the pump and blue the probe signal) (D) Topography overlaid with the difference of the two maps (E) Histograms of the PP-KPFM maps separated into grain and grain boundary responses.

At these frequencies we observed a ~1.8 µs relaxation and a ~1.3 µs generation time constant. Physically at this timescale we would expect the electric potential dynamics to be governed by charge carrier diffusion and recombination. Considering this, the asymmetry in the square response can be explained by effect of non-radiative recombination and/or charge carrier diffusion times[5]. After identifying the time constant of interest, we obtained pump-probe KPFM images of the CsFAMAPbBrI at the two different pulse positions indicated in the inset of **Figure 4B** and **C**. By subtraction of these maps we gain a high frequency electrostatic force image that is directly related to the $V_{cpd}$ change of the sample at the timescale of the applied pulses (150 kHz), representing measurement in the dark and under illumination. **Figure 4D** shows the topography overlaid with the calculated difference. The result of the subtraction suggests heterogeneity among grains and more importantly a larger change at the grain boundaries of the structures. Masking of the topography (**Figure S5**) and calculation of the grain and grain boundary histograms confirms this observation as can be seen in **Figure 4E**. In order to exclude effects of phase segregation during the time between taking the two images, we repeated the measurements with a lower illumination intensity of 1 sun (**Figure S6**). These measurements showed the same contrast, confirming that the observed effect is not due to light-induced phase segregation of perovskites.

We interpret these findings as a larger potential change due to a higher number of electron-hole recombination events at the grain boundaries over the course of ~6.7 μs for the applied 150 kHz pulse frequency. Presence of intrinsic lower bandgap iodide rich phases have been observed for mixed-halide Cs-incorporated perovskite [33,34]. Such an intrinsically segregated phase would provide smaller energy recombination for the charge carriers which could lead to faster recombination time. Furthermore, it has been shown that charge carriers preferentially localize in iodide-rich regions, which could lead to more radiative recombination taking place closer to the grain boundaries explaining the observed contrast in our results [35,36]. We note that the contrast in **Figure 4D** could also indicate a difference due to diffusion of charge carriers at the timescale of the applied pulses. This would suggest the diffusion of the generated carriers from the bulk of the grains towards the grain boundaries further confirming the electron accumulating effect of the GBs.

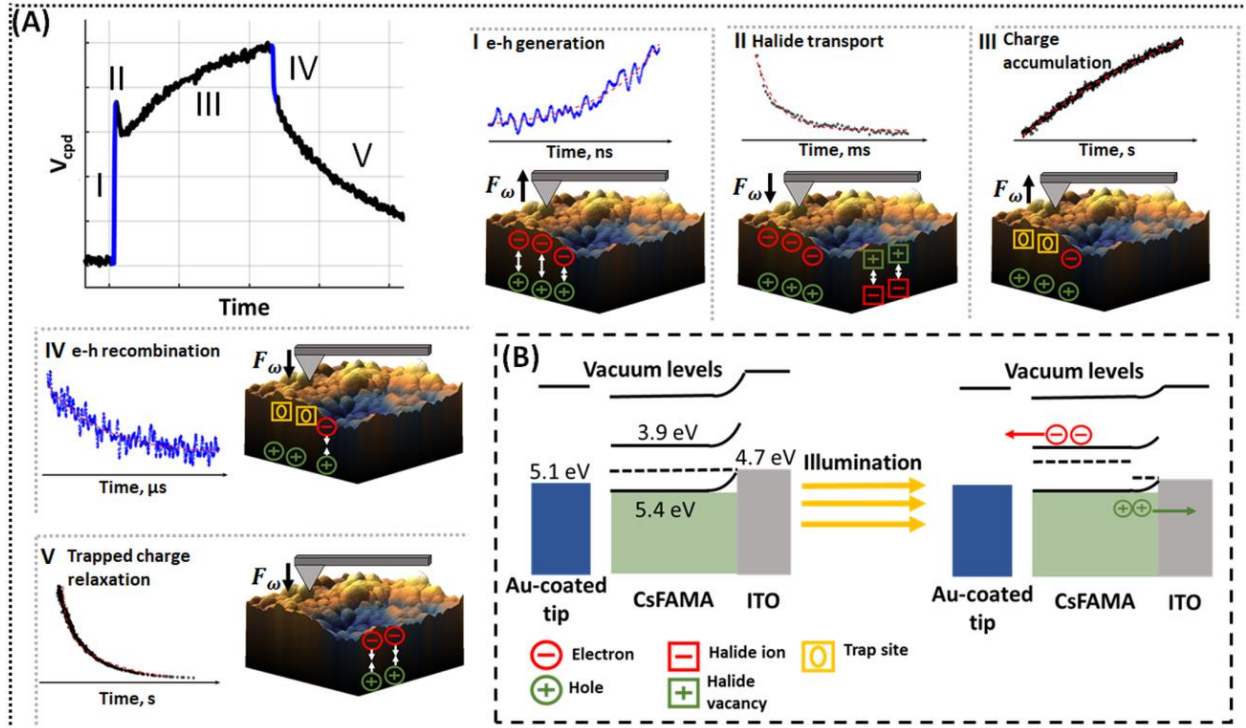

**Figure 5.** Detailed picture of electrostatic force response of mixed-halide CsFAMA. (A) Stepwise explanation of charge carrier transport from the electrostatic force response (B) Band diagram in dark and illuminated conditions.

**Collective model of CsFAMAPbBrI response**. The described effects can be summarized as follows. **Figure 5B** shows the band alignment of the structure. The work function of ITO is around 4.7 eV, the conduction and valence band edges of the perovskite were reported to be ~3.9 eV and ~5.4 eV, respectively, while the work function of the Au coated tip can be considered to be around 5.1 eV [16,37,38]. According to the band diagram, the generated electrons will preferentially move away from the perovskite-ITO interface upon illumination.

Thus, as shown in **Figure 5A** at step I we resolve movement of electrons towards the top interface confirmed by the $V_{cpd}$ shift in **Figure 3**. Intrinsically under dark conditions the work function of the GBs is larger, indicating a downward band bending that would promote local collection of electrons. Previous studies proposed intrinsic iodide rich regions in mixed-cation perovskites, which corroborate collection of photoinduced charge-carriers [33,35,36,39]. Pump-probe KPFM imaging indicated diffusion of charge carriers towards the GBs with faster recombination dynamics that can be due to the localized iodide-rich regions. Consequently, after the electron-hole pair separation and charge localisation at the GBs, negatively charged halide motion occurs away from the top surface in step II. Promotion of halide migration has been linked to higher charge densities [40]. At step III negative charge carriers accumulate at grain boundaries as shown in the $\Delta V_{cpd}$ map. The timescale of this effect indicates ionic origin. Redistribution of the local electric field and vacancy density can affect ionic motion [41]. Afterwards, turning off the illumination results in a more positive potential at this interface due to electron-hole recombination (step IV). Finally, through relaxation of the trapped charges the recorded potential reaches a more positive value close to the initial $V_{cpd}$ (step V).

**Application to multilayer structures.** After describing the response of the perovskite, we applied the same approach to multilayer structures to investigate the effect of transport layers. A thin TiOx layer on top of PCBM in an inverted configuration has shown more efficient charge extraction and improved optical properties [16]. The structure of the measured multilayer architectures can be found in the caption for **Figure 6**. First, in **Figure 6A** we show the electrostatic response to an illumination pulse for the different structures. Qualitative comparison of the responses shows that by depositing the perovskite on a hole-transport material, the previously described ionic motion (step II) and charge accumulation (step III) is amplified and accelerated. This can be explained by extraction of holes into the NiO$_x$, leaving excess electrons in the perovskite.

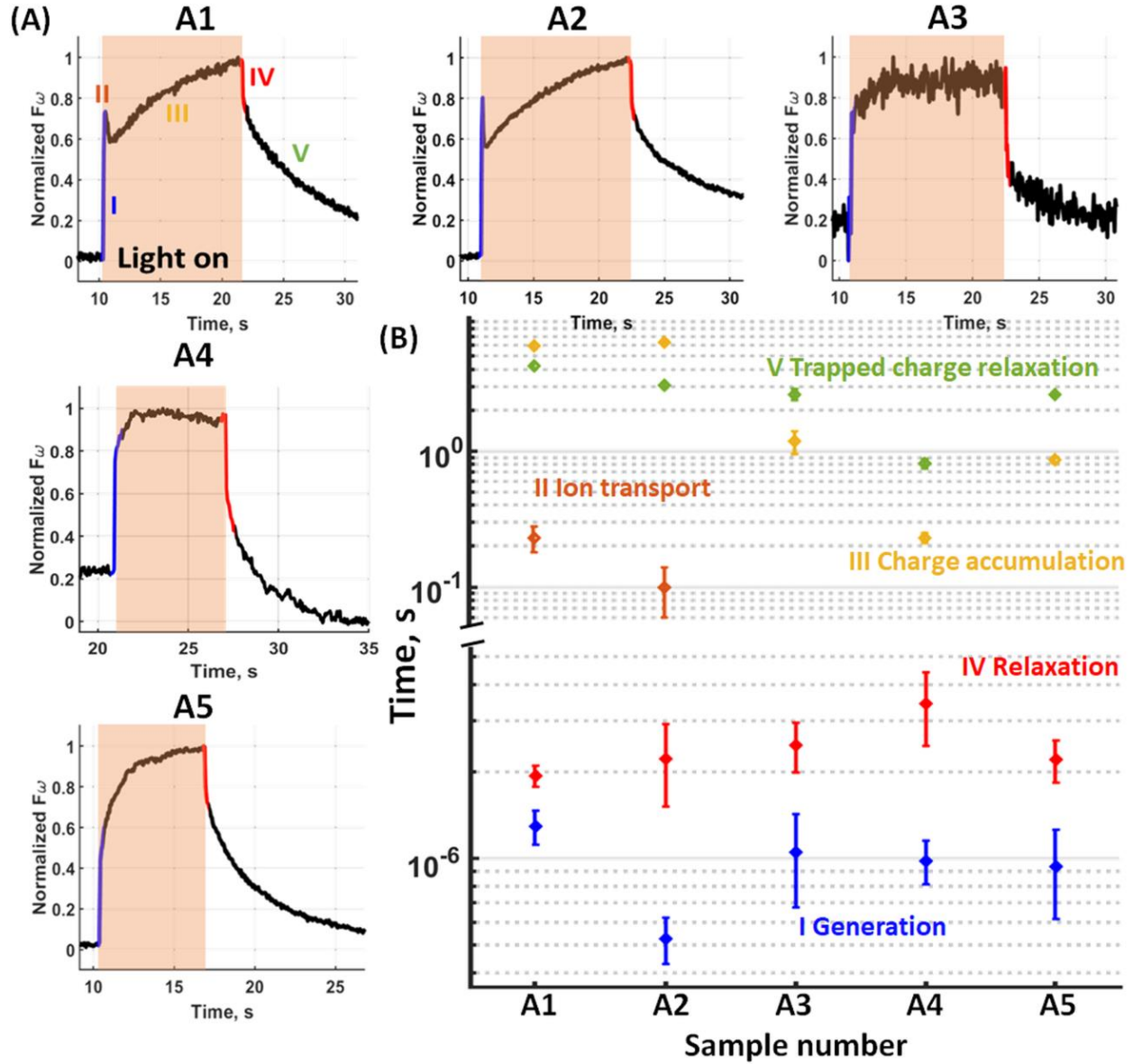

**Figure 6.** Carrier dynamics of multilayer CsFAMAPbBrI structures. (A) $F_\omega$ time response to a single light pulse for different multilayer structures. (B) Time constants extracted from each shown step and from pump probe responses for the different samples. The characterized samples were of the following structures on Glass/ITO: A1 Perovskite, A2 NiOx/Perovskite, A3 NiOx/Perovskite

/PCBM, A4 NiOx/Perovskite /PCBM/TiOx(10-15 nm), A5 NiOx/Perovskite /PCBM/TiOx(40-50 nm). The blue and red regions of the time traces correspond to the electron-hole pair generation and relaxation respectively.

A number of studies have shown a direct link between electron density and promotion of ion diffusion through electron-phonon coupling and halide vacancy polarization [42–46]. Considering passivation of halide vacancies via electrons, the absence of step II for structures with ETM on top (A3-A4-A5) thus corroborates its ionic origin. Additionally, the surface charge accumulation was accelerated as seen in the extracted fitting time constant in **Figure 6B** for step III and V. These observations are in line with findings that have shown passivation of deep traps by fullerene-halide radicals [1]. The extraction of the data points and uncertainty bars is detailed in the Methods section. **Table S1** contains the extracted parameters from the resulting monoexponential fits to the separate dynamical processes. The results indicate that the constraints such as charge accumulation or enhanced ion migration at the grain boundaries can be resolved by appropriate surface passivation [47]. Addition of a thin $TiO_x$ top layer results in a clear increase in the rate of charge accumulation and relaxation. On the other hand, a thicker $TiO_x$ layer decreases the rate indicating a reduced conductivity.

Application of the pump-probe KPFM approach confirms this observation. **Figure 6B** includes the extracted generation and relaxation times from single point time trace measurements. Extraction of the time constants is detailed in Materials and Methods. Clearly, in the cases of the perovskite top layer (A1, A2) we do not resolve significant differences in the fast charge carrier relaxation times. In cases with the ETMs on top of the active layer, the relaxation time increases indicating a longer charge carrier diffusion time and/or recombination time. This can be due to interfacial energy barriers or an increase in trap mediated recombination. This trend continues for the thin $TiO_x$ layer. However, the samples with a 40-50 nm thick $TiO_x$ layer showed a ~1 μs decrease in the relaxation time. This result indicates an increased rate of recombination or a decrease in the overall diffusion length of the charge carriers. As the macroscale IV and power conversion efficiency (PCE) measurements showed higher fill factor and PCE for the thin $TiO_x$ capping layer compared to the thick, we propose that this is due to a high density of formed trap sites in the imperfect top layer [16]. As access only to the top layer of the device limits the characterization of layer specific mechanism, in order to understand the complete nanoscale effect of the double ETM structure, further time resolved KPFM measurements on cross-sectional structures could provide a deeper understanding for the observed effects.

**Conclusion**

In conclusion, we applied KPFM techniques to characterize the charge carrier dynamics of mixed-cation mixed-halide perovskite ($Cs_{0.05}(FA_{0.83}MA_{0.17})_{0.95}PbI_{3-x}Br_x$) structures under light excitation. Conventional KPFM measurements allowed us to separate light-induced effects of different timescales such as the charge carrier lifetimes and trapped charge relaxations. KPFM mapping revealed charge accumulation and trapped charge relaxation at the grain boundaries with dynamics in the order of seconds ($\tau_{II} = 0.23 \pm 0.05$ s). Pump-probe implementation of open-loop KPFM revealed higher activity at the GBs indicating diffusion of charge-carriers towards the GBs and faster local recombination. From our results and the corresponding literature we outline a collective model for mixed-halide CsFAMA that encompasses the spatially distributed nanoscale electric effects. This model can facilitate targeted use of the mechanisms described for advanced optoelectronic applications. Furthermore, the nanoscale charge dynamics of multilayer structures were measured by KPFM, showing a direct correlation with macroscale electric and optical measurement results. Further investigation of these characteristics by advanced SPM techniques will direct further progress in perovskite fabrication as well as optimized transport layer preparation. Finally, we want to point out that the methodologies for nanoscale investigation of carrier dynamics employed here can be readily transferred to other research fields ranging from advanced optoelectronic applications of 2D materials to optically active molecules in life science applications [48,49].

**Acknowledgement**

David Toth and Filipe Richheimer acknowledge financial support from the European Union`s Horizon2020 research and Innovation programme under the Marie Sklodowska-Curie grant agreement No. 721874 (SPM2.0). Bekele Hailegnaw acknowledges the financial support of the Austrian Academy of Science in the framework of the Chemical Monthly Fellowship (MoChem), and Austrian Research Promotion Agency (FFG) under the project Flex!PV-2.0-85360. Fernando Castro and Sebastian Wood acknowledge financial support from the UK Department for Business, Energy and Industrial Strategy (BEIS) through the National Measurement System. Technical discussions with Ivan Alic are acknowledged. This work has been supported by EFRE Project IWB 2018, Nr. 98292, NMBP project MMAMA 761036 and FWF Project P28018-B27.

**Methods**

**AFM measurements.** Two separate AFM setups were used for KPFM characterization. Measurements presented in the Supporting Information were taken on a Keysight Technologies 5600 large stage microscope in ~1% relative humidity conditions. A modulatable CW 405 nm laser diode was used as the illumination source. Demodulation of the photodetector signal was done with the internal lock-in amplifiers of the AFM setup using 100 Hz bandwidth. 13 kHz resonance frequency (PtSi-CONT, Nanosensors) silicon AFM tips with PtSi coating were used for these measurements. Square wave and pulse waveform generation was done with a Keysight Technologies 33500b series waveform generator. Rising and falling edges of the generated pulses were 2.9 ns across the complete modulation depth. The PWM measurements were performed keeping a constant 100 nm distance above the surface. The experiments were performed in an amplitude modulated 2-pass KPFM manner.

Results presented in the main figures were taken on an AIST-NT Combiscope 1000 AFM system in nitrogen atmosphere. For excitation a 633 nm He-Ne laser was used modulated by a New Focus 4102NF electro-optical modulator. For these measurements 70 kHz nominal resonance frequency gold coated tips were employed. Imaging was done in lift mode keeping 10 nm distance with respect to the topography pass. We note that the AFM measurements reported included rigorous efforts to minimize and remove any light-induced topographical change. Additionally, positioning of the laser beam and application of band pass optical filters allowed us to avoid any crosstalk in the laser read-out signal due to the applied laser illumination.

**Photoluminescence measurements**. Spectrally resolved photoluminescence was measured using a Horiba Labram Evolution HR800 system. For optical excitation a 633 nm He-Ne laser was used. The collected signal was diffracted on a 300 lines/mm spectral grating.

**Fabrication of perovskite structures**. Glass substrates coated with indium tin oxide (ITO, 15 $\Omega/cm^2$), lead iodide ($PbI_2$, Sigma Aldrich, 99.9 %), lead bromide ($PbBr_2$, Sigma Aldrich, 99.99 %), [6,6]-phenyl-C61-butyric acid methyl ester (PCBM, Solenne BV), cesium iodide (CsI, Sigma Aldrich, 99.99 %), formamidine acetate salt ($HN=CHNH_2 \cdot CH_3COOH$, Sigma Aldrich, 99 %), hydroiodic acid (HI, 57 wt % in $H_2O$), hydrobromic acid (HI, 57 wt % in $H_2O$), methylamine ($CH_3NH_2$, Aldrich, 33 wt % in absolute ethanol), N,N-dimethylformamide (DMF, anhydrous,

Sigma Aldrich), dimethylsulfoxide (DMSO, Anal. R. VWR chemicals, 99.5 %) and chlorobenzene were used as purchased. Nickel chloride hexahydrate ($NiCl_2 \cdot 6H_2O$, Sigma Aldrich, 99.9 %) and sodium hydroxide (NaOH, Sigma Aldrich, > 98 %) were used to synthesize $NiO_x$ nanoparticles following the procedure reported by X. Yin et al. [50]. To synthesize $TiO_x$ sol-gel, titanium (IV) isopropoxide ($Ti[OCH(CH_3)_2]_4$, Sigma Aldrich, 99.9+ %), 2-methoxyethanol ($CH_3OCH_2CH_2OH$, Sigma Aldrich, 99.9 %), isopropanol and ethanolamine ($H_2NCH_2CH_2OH$, Sigma Aldrich, 99 %) were used. $TiO_x$ sol-gel and organic salts ($CH_3NH_3Br$ and $CH(NH_2)_2I$) were synthesized. following the same synthesis procedure mentioned before[16].

The five measured perovskite structures consisted of the perovskite directly deposited on glass substrates coated with indium tin oxide (ITO), the perovskite placed on glass/ITO/Nickel-oxide, a Glass/ITO/$NiO_x$/Perovskite/PCBM structure and Glass/ITO/NiOx/Perovskite/PCBM/TiOx with different thicknesses of titanium oxide from 10 to 50 nm. Here we refer to these as A1, A2, A3, A4, A5 correspondingly.

To prepare the samples, glass substrates coated with indium tin oxide (ITO) were sequentially cleaned in acetone, detergent, deionized water and isopropanol (IPA). $NiO_x$ particles (20 mg/mL in water) were cracked using an ultrasound sonicator (UP50H, 50 Watt, frequency 30 kHz), followed by filtration through 0.45 μm pore-size polytetrafluoroethylene (PTFE) syringe filter. The filtrate dispersion was spin-coated on the ITO-coated substrate at 4000 rpm for 30 s. Then the film was annealed at 140 °C for 20 min. For the deposition of perovskite films and further processing the substrates were transferred into a glove box.

Triple-cation mixed-halide perovskite ($Cs_{0.05}(FA_{0.83}MA_{0.17})_{0.95}PbI_{3-x}Br_x$) solution was prepared by dissolving $PbI_2$ (507.5 mg), $CH(NH_2)_2I$ (172 mg), $PbBr_2$ (73.5 mg) and $CH_3NH_3Br$ (22.4 mg) in a mixture of N,N-dimethylformamide (DMF) and dimethyl sulfoxide (DMSO) solvents (4:1 (v/v) ratio), followed by stirring at 45 °C. Then, 5 % (v/v) of CsI (1.5 M in DMSO) was added into the mixture and stirred for about 3 hrs. The perovskite solution was deposited on top of ITO substrates to prepare S1 and ITO/$NiO_x$ films to fabricate S2, S3, S4 and S5. Spin-coating was conducted in two-steps at 1500 rpm for 10 s with ramp: 150 rpm/s and at 6000 rpm for 30 s with ramp: 3000 rpm/s, with in-situ anti-solvent quenching by dropping about 0.15 mL of chlorobenzene for about 3 s. The films were annealed at 100 °C for 1 hr. Then films were cooled to room temperature. To prepare S3, S4 and S5, PCBM dissolved in a mixture of chlorobenzene and chloroform (50:50 volume ratio, 2 wt%) was spin-coated on top of the perovskite films. Diluted $TiO_x$ sol-gel solution was spin-coated on top of PCBM at 5000 rpm for 30 s for S4 and at 3000 rpm for 30 s to finish S5, followed by annealing at 110 °C for about 5 min in ambient air.

**Extraction of time constants**. A custom Wolfram Mathematica script was used for the extraction of time constants from the pump-probe responses utilizing a Heaviside function with exponential rising and falling edges. The defined function was fitted to 6 periods of the recorded signal. In Figure 7 mean value of the resulting corrected time constant is presented with the standard deviation. Time constants from the low frequency single light pulse responses were extracted from monoexponential fitting in Origin. Data shown in Figure 7 represents the extracted time constants with the error resulting from the fit.

**Supporting Information**

Photoluminescence response of the mixed halide CsFAMA perovskite, first and second harmonic of the electrostatic force response to an illumination pulse, mask generated from the topographical data shown in Figure 3, a recorded pump-probe KPFM time trace with variation of extracted time constant versus probing pulse width, mask generated from topographical data for Figure 5, results of pump-probe open-loop Kelvin probe force microscopy measured under 633 nm ~1 sun illumination intensity, table including exponential fit parameters extracted from single light pulse responses.

For Table of Contents Only

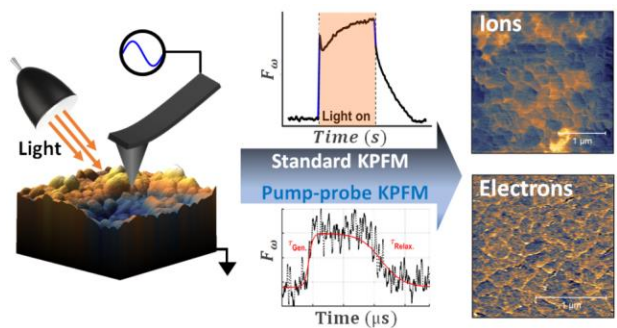